\documentclass[pre,preprint,superscriptaddress]{revtex4}
\usepackage{epsfig}
\usepackage{latexsym}
\usepackage{amsmath}
\begin {document}
\title {Dynamical properties of the synchronization transition}
\author{Michel Droz}
\affiliation{Department of Physics, University of Geneva, CH 1211
Geneva 4, Switzerland}
\author{Adam Lipowski}
\affiliation{Department of Physics, University of Geneva, CH 1211
Geneva 4, Switzerland}
\affiliation{Department of Physics, A.~Mickiewicz University,
61-614 Pozna\'{n}, Poland}
 %%%%%%%%%%%%%%%%%%%%%%%%%%%%%%%%%%%%%%%%%%%%%%%%%%%%%%%%%%%%%%%%%%%%%%%%%%%%%%%
\begin {abstract}
We study the dynamics of the synchronization transition (ST) of
one-dimensional coupled map lattices.  For the Bernoulli map it was
recently found by Ahlers and Pikovsky (Phys.~Rev.~Lett.~{\bf 88},
254101 (2002)) that the ST belongs to the directed percolation (DP)
universality class.  Spreading dynamics confirms such an
identification, only for a certain class of synchronized
configurations.  For homogeneous configurations spreading exponents
$\eta$ and $\delta$ are different than DP exponents but their sum
equals to the corresponding sum of DP exponents.  Such a relation is
typical to some models with infinitely many absorbing states.
Moreover, we calculate the spreading exponents for the tent map for
which the ST belongs to the bounded Kardar-Parisi-Zheng (BKPZ)
universality class.  Our estimation of spreading exponents are
consistent with the hyperscaling relation.  Finally, we examine the
asymmetric tent map.  For small asymmetry the ST remains of the
BKPZ type.  However, for large asymmetry a different critical
behaviour appears with exponents being relatively close to the ones of DP.
\end{abstract}
\maketitle
%%%%%%%%%%%%%%%%%%%%%%%%%%%%%%%%%%%%%%%%%%%%%%%%%%%%%%%%%%%%%%%%%%%%%%%%%%%%%
\section{introduction}
Recently, synchronization of chaotic systems received considerable
attention~\cite{FUJISAKA}.  These studies are partially motivated by
experimental realizations in lasers, electronic circuits, and chemical
reactions~\cite{EXP}.  An interesting problem concerns synchronization
in spatially extended systems~\cite{EXTENDED}.  It turns out that in
such systems synchronization can be regarded as a nonequilibrium phase
transitions.  The determination of the universality classes for
nonequilibrium phase transitions is a problem much debated in the
literature. Thus it is natural to ask whether the
synchronization transition (ST) can be incorporated into an already
known universality class.  For certain cellular automata the
synchronized state is actually an absorbing state of the dynamics, and
ST in such a case was found to belong to the Directed Percolation (DP)
universality class~\cite{GRASS1999}.  This problem is more subtle for
continuous chaotic systems as e.g., coupled map lattices
(CML)~\cite{KANEKO}.  For CML a perfect synchronized state is never
reached in finite time which weakens the analogy with DP.  Recently
Pikovsky and Kurths argued that synchronization of CML should
generically be described by the so-called bounded Kardar-Parisi-Zheng
(BKPZ) universality class~\cite{PIKKURTHS,TU}.

However, it  was observed that for noise-induced
synchronization~\cite{LIVI} as well as for a system of two interacting
identical CML~\cite{AHLERS}, the synchronization transition belongs to
the DP universality class.  It then was argued that DP critical
behaviour might emerge for systems with discontinuous local maps, for
which a linear approximation, that leads to the BKPZ universality
class, breaks down~\cite{AHLERS}.

It is well known that  models with a single absorbing state exhibit a
DP criticality~\cite{GRASJAN}.  At first sight one can think that the
synchronized state is unique and that this is the reason of its
relation with DP criticality in some cases.  However, if one studies
the dynamical properties of the ST by means of the so-called spreading
approach~\cite{SPREADING}, several  synchronized states could be
considered and some dynamical properties depend on the choice of a
synchronized state. For example, the spreading exponents which we
measure for homogeneous synchronized state take non-DP values.
However, these exponents satisfy a more generalized scaling relation
that also holds for some models with infinitely many absorbing
states~\cite{INFINITY}.

In addition to that, we measured spreading exponents for ST belonging
to the BKPZ universality class.  Our results, obtained for the
symmetric tent map, together with previous estimations of other
exponents in this universality class, show that the hyperscaling
relation is satisfied in this case.  Finally, we examine the asymmetric
tent map.  For not too large asymmetricity the ST remains of BKPZ
type.  However, for large asymmetricity the nature of the ST changes
and the critical exponents that we measured are relatively close to
that of the DP. It shows that the non-BKPZ behaviour appears even for
continuous maps.

In section II we define the model and briefly describe the simulation method.
The results are presented in section III and a final discussion is given  in section IV.
%%%%%%%%%%%%%%%%%%%%%%%%%%%%%%%%%%%%%%%%%%%%%%%%%%%%%%%%%%%%%%%%%%%%%%%%%%%
%%%%%%%%%%%%%%%%%%%%%%%%%%%%%%%%%%%%%%%%%%%%%%%%%%%%%%%%%%%%%%%%%%%%%%%%%%%
\section{Model and simulation method}
Our model is the same as the one examined recently by Ahlers and Pikovsky 
(AP) and 
consists of two coupled CML's~\cite{AHLERS},
\begin{equation}
\begin{pmatrix} u_1(x,t+1) \\ u_2(x ,t+1)\end{pmatrix} =
\begin{pmatrix} 1-\gamma & \gamma \\ \gamma & 1-\gamma\end{pmatrix}\times
\begin{pmatrix} (1+\epsilon\Delta)f(u_1(x,t+1)) \\
(1+\epsilon\Delta)f(u_2(x,t+1))\end{pmatrix},
\label{model}
\end{equation}
$\Delta v$ the discrete Laplacian $\Delta v(x)=v(x-1)-2v(x)+v(x+1)$.
Both space and time are discretized, $x=1,2\ldots,L$ and $t=0,1,\ldots$.
Periodic boundary conditions are imposed $u_{1,2}(x+L,t)=u_{1,2}(x,t)$ and
similarly to previous studies we set the intrachain coupling $\epsilon=1/3$.
Varying the interchain coupling $\gamma$ allows us to study the transition between synchronized and chaotic phases.
Local dynamics is specified through a nonlinear function $f(u)$ and several cases will be discussed below.

We introduce a synchronization error $w(x,t)=|u_1(x,t)-u_2(x,t)|$ and
its spatial average $w(t)=\frac{1}{L}\sum_{x=1}^L w(x,t)$.
The time average of $w(t)$ in the steady state will be simply denoted as $w$.
In the chaotic phase, realized for sufficiently small $\gamma < \gamma_c$, one has $w>0$, while in the synchronized phase ($\gamma > \gamma_c$), $w=0$.
Moreover, at criticality, i.e. for $\gamma = \gamma_c$, $w(t)$ is expected to have a power-law decay to zero $w(t)\sim t^{-\Theta}$.

It is well known that the spreading dynamics is a very effective method
to study phase transitions in models with absorbing
states~\cite{SPREADING,HAYE}.  In this method one prepares the model in
an absorbing state and then one locally sets the activity and monitor its
subsequent evolution.  Typical observables in this method are the
number of active sites $N(t)$, the probability $P(t)$ that activity
survives at least up to time $t$ and the average square spread of
active sites $R^2(t)$.  However, as we already mentioned, our model
never reach a perfectly synchronized state in a finite time.  To define
$P(t)$ we have to introduce a small threshold $p$ and consider a system
as "perfectly" synchronized when $w(t)<p$.  As we will show below
estimation of the critical value of the interchain coupling $\gamma_c$
as well as some exponents seems to be independent on the precise value
of $p$, as long as it remains small.

As for the other observables, we define them in the following, $p$-independent
way
\begin{equation}
N(t)=w(t), \quad R^2(t)=\frac{1}{w(t)}\sum_x w(x,t)(x-x_0)^2,
\label{observ}
\end{equation}
where $x_0$ denotes the site where the activity was set initially.

According to general scaling arguments~\cite{SPREADING}, we expect that at
criticality (i.e., for $\gamma=\gamma_c$):
$w(t)\sim t^{\eta}$, $P(t)\sim t^{-\delta}$, and $R^2(t)\sim t^{z}$. These relations define the critical exponents $\eta$, $\delta$, and $z$.

Note that in general the critical behavior of $w(t) \sim t^{\eta}$ in the
spreading case might be different from the non-spreading one
$w(t)\sim t^{-\Theta}$ discussed above.

%%%%%%%%%%%%%%%%%%%%%%%%%%%%%%%%%%%%%%%%%%%%%%%%%%%%%%%%%%%%%%%%%%%%%%%%%%
%%%%%%%%%%%%%%%%%%%%%%%%%%%%%%%%%%%%%%%%%%%%%%%%%%%%%%%%%%%%%%%%%%%%%%%%%%
\section{Results}
%%%%%%%%%%%%%%%%%%%%%%%%%%%%%%%%%%%%%%%%%%%%%%%%%%%%%%%%%%%%%%%%%%%%%%%%%%
\subsection{Bernoulli map}
First, let us consider the case when $f(u)$ is a Bernoulli map, namely
$f(u)=2u({\rm mod}~1)$ and $0\leq u\leq 1$.  Measuring $w$ in the
steady state and its time dependence $w(t)$, AP concluded that the
synchronization transition in this case takes place at
$\gamma=\gamma_c=0.2875(1)$ and belongs to the DP universality class.

\begin{figure}
\centerline{\epsfxsize=8cm
\epsfbox{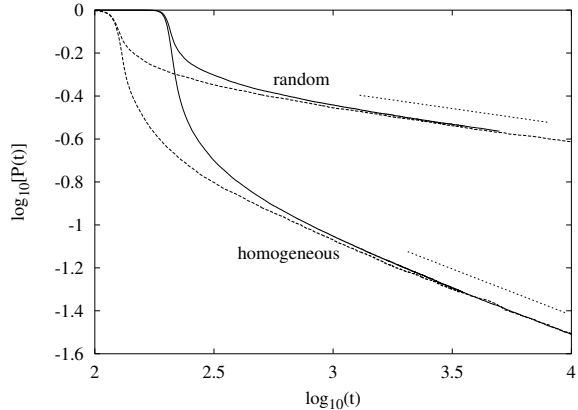}
}
%\figspace
\caption{
The survival probability $P(t)$ for the Bernoulli map at
$\gamma=\gamma_c=0.2875$ and for
random and homogeneous absorbing states as a function of time $t$.
Calculations were done for the threshold $p=10^{-10}$ (dashed line) and
$10^{-16}$ (solid line).
The upper straight dotted line have a slope which corresponds to the DP value
$\delta=0.1595$ and for the lower line $\delta=0.43$.
}
\label{time2875a}
\end{figure}

\begin{figure}
\centerline{\epsfxsize=8cm
\epsfbox{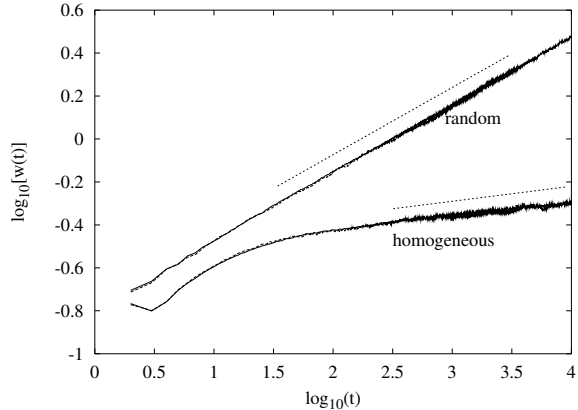}
}
%\figspace
\caption{
The averaged difference $w(t)$ as a function of time $t$
for the Bernoulli map at $\gamma=\gamma_c=0.2875$ and for random and
homogeneous absorbing states.
Calculations were done for
the threshold $p=10^{-10}$ (dashed line) and
$10^{-16}$ (solid line).
The upper straight dotted line have a slope which corresponds to the DP value
$\eta=0.3134$ and for the lower straight line $\eta=0.05$.
}
\label{time2875}
\end{figure}

\begin{figure}
\centerline{\epsfxsize=8cm
\epsfbox{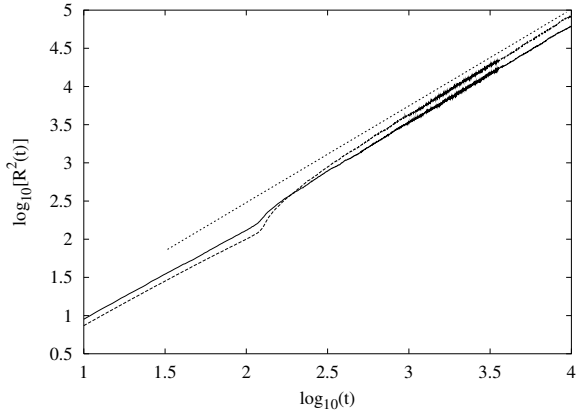}
}
%\figspace
\caption{
The average spread square $R^2(t)$ as a function of time $t$
for the Bernoulli map at $\gamma=\gamma_c=0.2875$ with random (solid line)
and homogeneous (dashed line) absorbing states.
The straight dotted line have a slope which corresponds to the DP value
$z=1.2652$.
}
\label{zdp}
\end{figure}

Our purpose here is to apply the spreading dynamics to this map.
First, we have to set the model in a synchronized state i.e., the variables of both
chains of CML's must take the same value.
But despite that constraint there is a considerable freedom in doing that.
Below we present results for two choices:\\
(i) Random synchronized state\\
In this case each pair of local variables takes a different
random value, i.e., $u_1(x,0)=u_2(x,0)=r(x)$, where $0\leq r(x)\leq 1$ and
$r(x)$ varies
from site to site.\\
(ii) Homogeneous synchronized state\\
In this case $r(x)=r$ and is constant for each $x$.\\
Having set the model in the synchronized state we initiate an activity
assigning
at a randomly chosen site $x_0$ a new random number either for $u_1(x_0)$ or
for $u_2(x_0)$.
Then we monitor the time evolution, measuring $w(t)$.
We always used the system size $L=2t_m+1$ where $t_m$ is the maximum simulation
time. In such a way the spreading is never affected by finite size effects.
Moreover, the data were  averaged over $10^3-10^4$ independent runs.
Final results shown in Fig.~\ref{time2875a}~-~Fig.~\ref{zdp} are obtained
for $\gamma=\gamma_c=0.2875$. 
Off-criticality the data deviate from the power law behaviour in a typical way.
%%%%%-------------------------------------
%%%---------------------------------------
Estimating the asymptotic slope of our data we conclude that for the
random synchronized state the exponents are in a good agreement with
very accurately known DP values:  $\delta_{DP}=0.1595$,
$\eta_{DP}=0.3137$, and $z=1.2652$~\cite{JENSEN}.  However, in the case
of homogeneous synchronized states, the exponents $\eta$ and
$\delta$ are clearly different and we estimate $\delta=0.43(1)$ and
$\eta=0.05(1)$.

Let us notice that a very similar situation occurs in certain models
with infinitely many absorbing states, where these critical exponents
are also found to depend on the choice of an absorbing
state~\cite{INFINITY}.  Although the spreading exponents are
nonuniversal in this class of models, they obey the following relation
\begin{equation}
\eta+\delta=\eta_{DP}+\delta_{DP}.
\label{scaling}
\end{equation}
It turns out that the estimated exponents for homogeneous synchronized
states also obey this relation within an estimated error.
From Eq.~(\ref{scaling}) and the hyperscaling relation~\cite{HAYE}
\begin{equation}
\eta+\delta+\Theta=\frac{dz}{2} \quad (d=1)
\label{hyper}
\end{equation}
it follows that the exponent $z$ must be constant and thus independent on the
choice of the absorbing state.
Indeed, in Fig.~\ref{zdp} one can see that both for the random and homogeneous
absorbing states the asymptotic increase of $R^2(t)$ is described by the same
exponent.

The main conclusion which follows from the above calculations is that
synchronization should be considered as a transition in a model with multiple
absorbing states rather than a single one.
Except for non-DP values of $\eta$ and $\delta$ it has essentially no other
consequences in the one-dimensional case.
For example all steady-state exponents keep the DP values.
However, the situation might be different in higher-dimensional models with
infinitely many absorbing states.
In particular, there are some analytical and numerical indications that
in such a case a non-DP criticality might appear~\cite{FRED}.\\
%%%%%-------------------------------------
%%%---------------------------------------
%%%%%-------------------------------------
%%%---------------------------------------
%%%%%%%%%%%%%%%%%%%%%%%%%%%%%%%%%%%%%%%%%%%%%%%%%%%%%%%%%%%%%%%%%%%%%%%%%
\subsection{Symmetric Tent map}
The second map examined by AP is the symmetric tent map defined as
$f(u)=1-2|u-1/2|, \quad (0\leq u\leq 1)$.
In this case they found that the synchronization transition belongs to the
BKPZ universality class and occurs at $\gamma=\gamma_c=0.17614(1)$.
A qualitative difference between the critical behaviour in the case of
Bernoulli and tent maps is attributed to the strong nonlinearity
(i.e., discontinuity) of the first map.
Another interesting feature reported by AP is a different behaviour of the
transverse Lyapunov exponent $\lambda_{\perp}$.
For the tent map $\lambda_{\perp}$ vanishes exactly at $\gamma_c$, while for the
Bernoulli map it vanishes inside a chaotic phase.

\begin{figure}
\centerline{\epsfxsize=8cm
\epsfbox{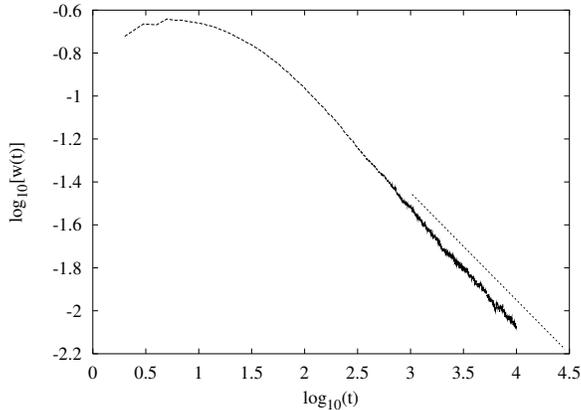}
}
%\figspace
\caption{
The averaged difference $w(t)$ as a function of time $t$
for the symmetric tent map at $\gamma=\gamma_c=0.17614$
and for random absorbing states.
Calculations were done for the threshold $10^{-10}$.
The least-square fit to the last decade date give $\eta=-0.53(5)$.
The dotted straight line has a slope which corresponds to $\eta=-0.5$.
}
\label{time17614}
\end{figure}

For the tent map we also performed the spreading-dynamics calculations
but only for random synchronized states.  Our results for the time
dependence of $w(t)$ are shown in Fig.~\ref{time17614}.  The asymptotic
decay of $w(t)$ is described by the exponent $\eta=-0.53(5)$ (note a
minus sign).

\begin{figure}
\centerline{\epsfxsize=8cm
\epsfbox{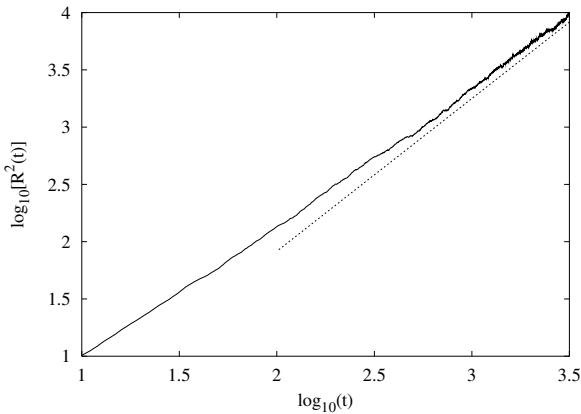}
}
%\figspace
\caption{
The average spread square $R^2(t)$ as a function of time $t$
for the symmetric tent map at $\gamma=\gamma_c=0.17614$ and random
absorbing states.
The straight dotted line have a slope which corresponds to the KPZ value
$z=1.3333$.
}
\label{rsq17614}
\end{figure}

As for the estimation of $\delta$ our results strongly
suggest that in this case $\delta=0$.
Indeed, in the examined range of the threshold $p$ we observed that all runs
survived until the maximum measured time $t_m=10^4$.
From the behaviour of $R^2(t)$ (Fig.~\ref{rsq17614}) we estimate $z=1.30(5)$,
which is in a good agreement with the expected KPZ value
$z=\frac{4}{3}$~\cite{TU,HWA}.

The hyperscaling relation provides a constraint on the exponents
$\eta,\ \delta,\ \Theta$, and $z$.
Since for the BKPZ universality class $z$ is a simple number (4/3), it is
tempting to assume that other exponents are also simple numbers.
Thus we might speculate that in this model the exact value of
$\eta$ is -0.5.
Then, from the hyperscaling relation (\ref{hyper}) together with $\delta=0$
and $z=\frac{4}{3}$ we obtain $\Theta=\frac{7}{6}=1.166\ldots$.
Such a value might be compared with numerical estimations of this exponent
which range from 1.1(1)~\cite{TU,HWA} to 1.26(3)~\cite{AHLERS}.
%%%%%-------------------------------------
%%%---------------------------------------
%%%%%-------------------------------------
%%%---------------------------------------
%%%%%%%%%%%%%%%%%%%%%%%%%%%%%%%%%%%%%%%%%%%%%%%%%%%%%%%%%%%%%%%
\subsection{Asymmetric Tent map}
In this subsection we examine the asymmetric tent map defined as
\begin{equation}
f(u)=\left\{\begin{array}{ll}
au & {\rm for}~0\leq u<1/a \\
a(1-u)/(a-1) & {\rm for}~1/a\leq u\leq 1, \\
\end{array}
\right.
\label{atent}
\end{equation}
and $1<a\leq 2$.
For $a=2$ this is the symmetric tent map.
Let us notice that in the limit $a\rightarrow 1$, the slope of the second part of this map
diverges.
Below we present the results of our calculations for $a=1.1,\ 1.02$ and 1.01.
\subsubsection{a=1.1}
\begin{figure}
\centerline{\epsfxsize=8cm
\epsfbox{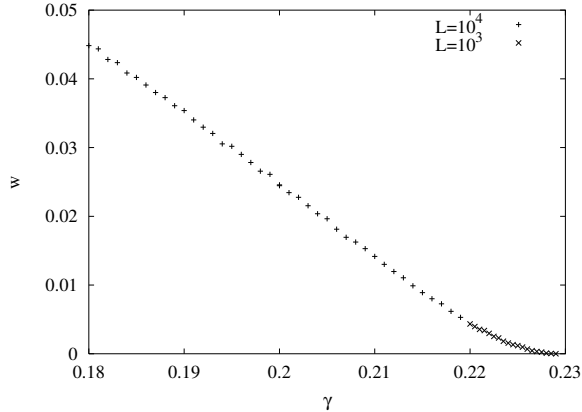}
}
%\figspace
\caption{
The steady-state synchronization error $w$ as a function of $\gamma$
for the asymmetric tent map ($a=1.1$).  The measurement was made
during  $t=10^5$ steps after $5\cdot 10^4$ steps of relaxation.  In the
range ($0.22 \le \gamma \le \gamma_c=0.2288$) $w$ is well fitted with a
power-law function $a(\gamma_c-\gamma)^{\beta}$, where $\beta=1.5$.}
\label{ro11}
\end{figure}

\begin{figure}
\centerline{\epsfxsize=8cm
\epsfbox{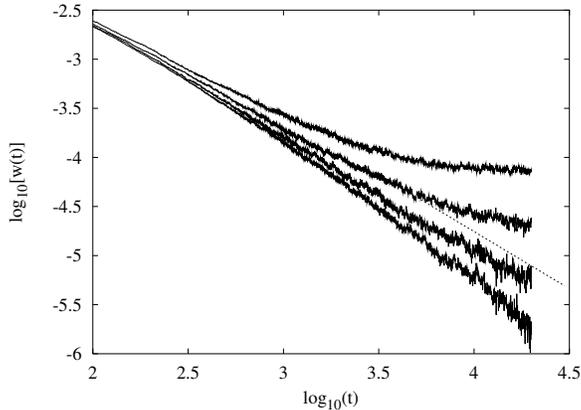}
}
%\figspace
\caption{
The time dependence of $w$
for the asymmetric tent map ($a=1.1$) and $\gamma$ equal to (from top)
0.228, 0.2285, 0.2288 (critical), and 0.229.
Calculations were done for $L=3\cdot 10^{5}$.
The dotted straight line has a slope corresponding to $\Theta=1.16$.
}
\label{time11}
\end{figure}

When $a$ is not too close to 1, model (\ref{model}) remains in the BKPZ
universality class.
In particular, for $a=1.1$ the estimated exponents $\beta=1.5(1)$
(Fig.~\ref{ro11}) and $\Theta=1.16(10)$ (Fig.~\ref{time11}) are in good
agreement with other estimations for this universality class.
%%%%%-------------------------------------
%%%---------------------------------------
%%%%%-------------------------------------
%%%%%-------------------------------------
\subsubsection{a=1.02}
\begin{figure}
\centerline{\epsfxsize=8cm
\epsfbox{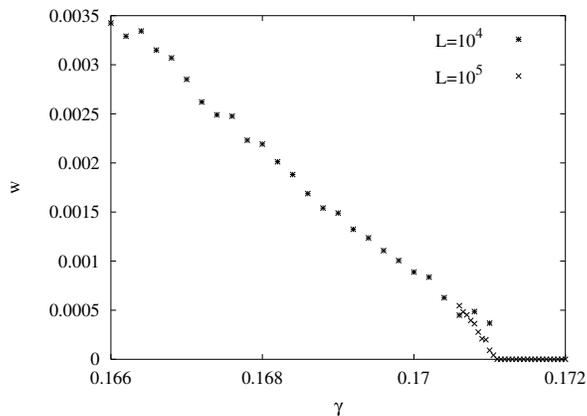}
}
%\figspace
\caption{
The steady-state synchronization error $w$ as a function of $\gamma$
for the asymmetric tent map ($a=1.02$).
The measurement was made during  $t=10^5$ steps after $5\cdot 10^4$ steps of
relaxation.
}
\label{rokpz102}
\end{figure}

\begin{figure}
\centerline{\epsfxsize=8cm
\epsfbox{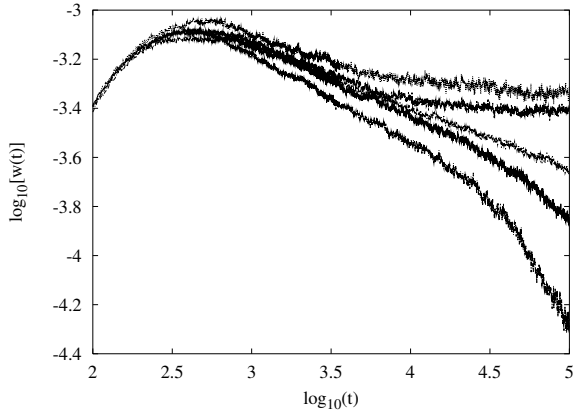}
}
%\figspace
\caption{
The time dependence of $w$
for the asymmetric tent map ($a=1.02$) and $\gamma$ equal to (from top)
0.1707, 0.1709, 0.17095, 0.1710, and 0.1712.
Calculations were done for $L=5\cdot 10^{4}$.
We identify the central curve as critical.
}
\label{time102}
\end{figure}

\begin{figure}
\centerline{\epsfxsize=8cm
\epsfbox{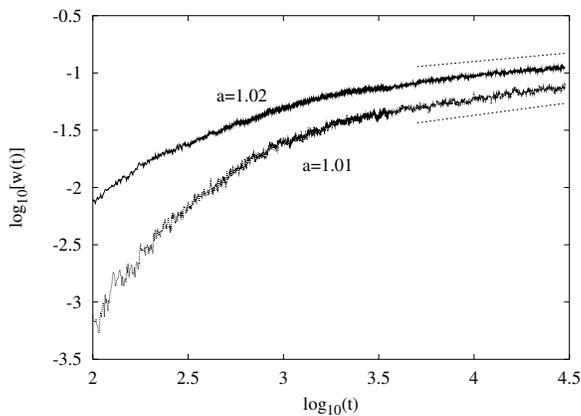}
}
%\figspace
\caption{
The averaged difference $w(t)$ as a function of time $t$
for the asymmetric tent map
and for random absorbing states.
Calculations were done for
$a=1.02$ with $ \gamma=\gamma_c=0.17095$ and for
$a=1.01$ with $ \gamma=\gamma_c=0.13695$.
The dotted straight lines have a slope which corresponds to $\eta=0.15$
(upper) and $\eta=0.22$ (lower).
}
\label{eta102}
\end{figure}

\begin{figure}
\centerline{\epsfxsize=8cm
\epsfbox{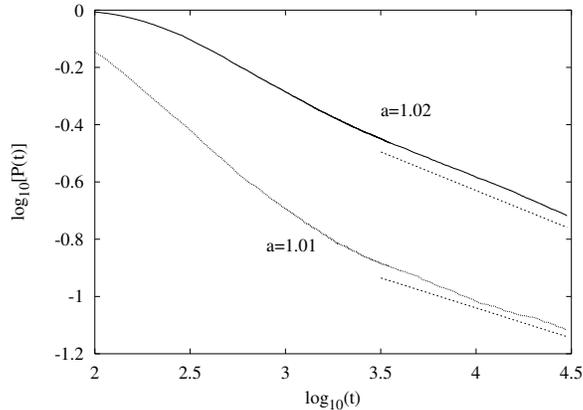}
}
%\figspace
\caption{
The survival probability $P(t)$ for the asymmetric tent map with
random absorbing states.
Calculations were done for
$a=1.02$ with $ \gamma=\gamma_c=0.17095$ and for
$a=1.01$ with $ \gamma=\gamma_c=0.13695$.
The dotted straight lines have a slope which corresponds to $\Theta=0.27$
(upper) and $\Theta=0.21$ (lower).
}
\label{teta102}
\end{figure}

\begin{figure}
\centerline{\epsfxsize=8cm
\epsfbox{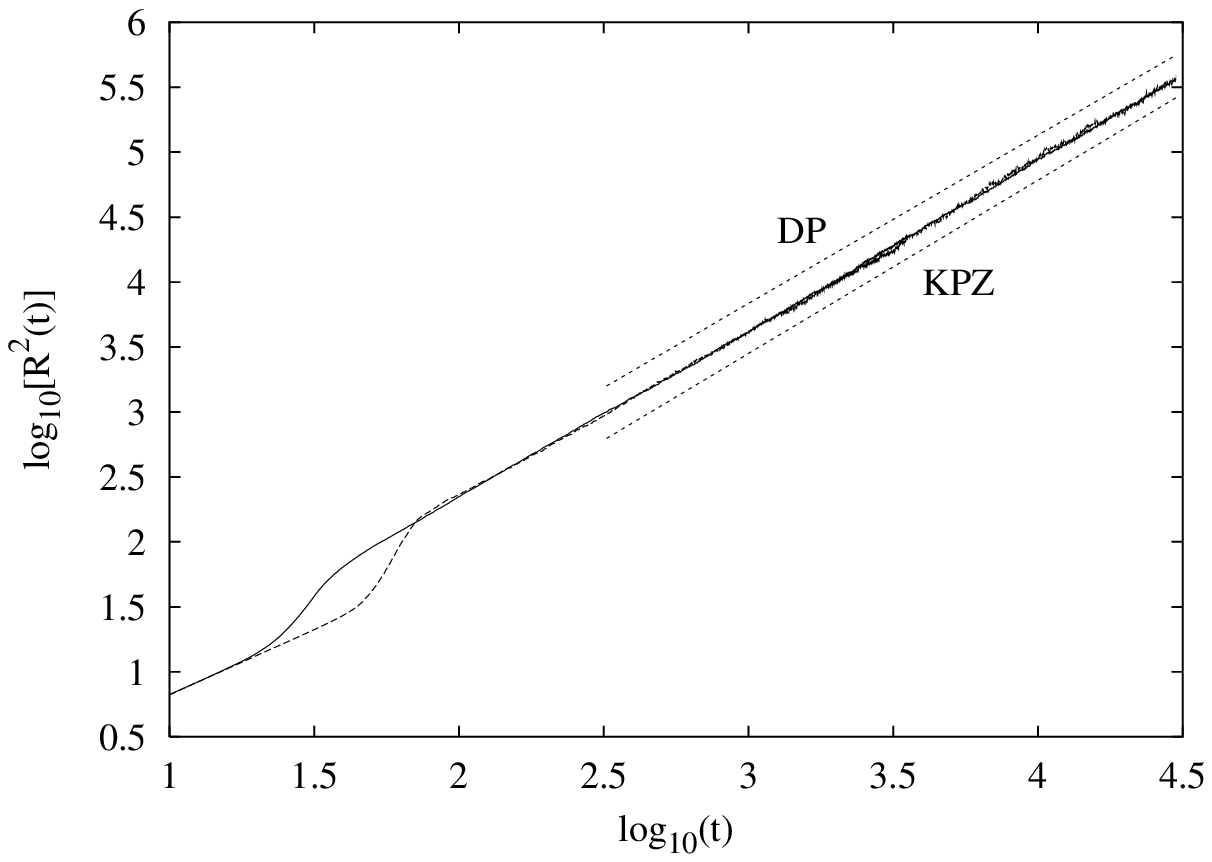}
}
%\figspace
\caption{
The average spread square $R^2(t)$ as a function of time $t$
for the asymmetric tent map and random absorbing states.
Calculations were done for
$a=1.02,\ \gamma=\gamma_c=0.17095$ (solid line) and
$a=1.01,\ \gamma=\gamma_c=0.13695$ (dashed line).
The straight dotted lines have slopes that correspond to DP and KPZ value of $z$.
}
\label{rsq102}
\end{figure}

However, for $a$ close to 1 the nature of the transition changes.
From the measurements of $w$ (Fig.~\ref{rokpz102}) and its time dependence
$w(t)$ (Fig.~\ref{time102}) we estimate that in this case $\gamma_c=0.17095$
and $\Theta=0.25(5)$.
Using the spreading dynamics calculations for the random synchronized
states at the critical point we estimate (Fig.~\ref{eta102}~-~Fig.~\ref{rsq102}):
$\eta=0.15(3),\ \delta=0.27(2)$, and $z=1.32(3)$.
All of the exponents, except $z$ are far from the BKPZ values.
%%%%%-------------------------------------
%%%---------------------------------------
%%%%%-------------------------------------
%%%%%-------------------------------------
%%%%%-------------------------------------
%%%%%-------------------------------------
%%%%%-------------------------------------
%============================================================================
%%%%%-------------------------------------
\subsubsection{a=1.01}
\begin{figure}
\centerline{\epsfxsize=8cm
\epsfbox{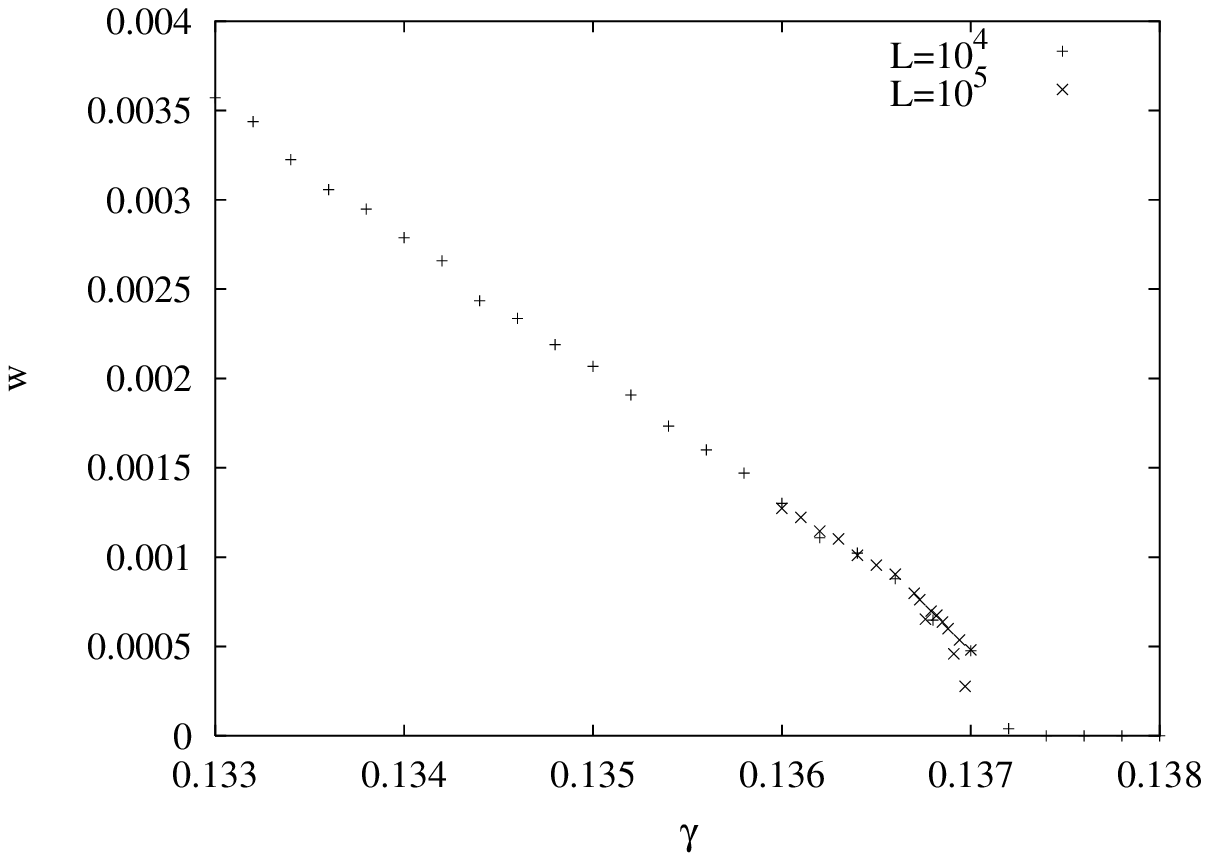}
}
%\figspace
\caption{
The steady-state synchronization error $w$ as a function of $\gamma$
for the asymmetric tent map ($a=1.01$).
The measurement was made during  $t=10^5$ steps after $5\cdot 10^4$ steps of
relaxation.
}
\label{ro101}
\end{figure}

\begin{figure}
\centerline{\epsfxsize=8cm
\epsfbox{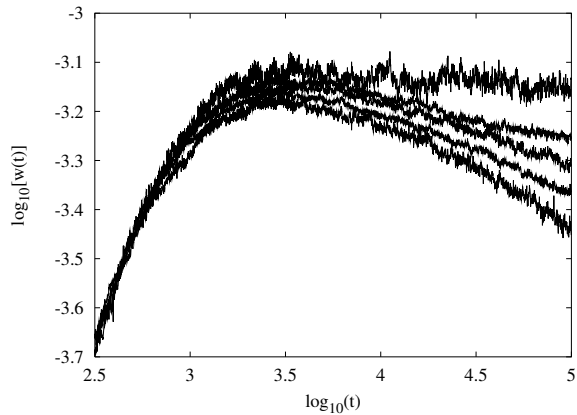}
}
%\figspace
\caption{
The time dependence of $w(t)$
for the asymmetric tent map ($a=1.01$) and $\gamma$ equal to (from top)
0.1368, 0.13685, 0.1369, 0.13695, and 0.137.
Calculations were done for $L=5\cdot 10^{4}$.
}
\label{time101}
\end{figure}

From the measurements of $w$ (Fig.~\ref{ro101}) and its time dependence
$w(t)$ (Fig.~\ref{time101}) we estimate that in this case $\gamma_c=0.13695$
and $\Theta=0.15(2)$.
Similarly to the $a=1.02$ our results for $w$ close to the critical point are
not sufficiently accurate to determine $\beta$.
Using the spreading dynamics calculations for the random synchronized
states at the critical point we estimate (Fig.~\ref{eta102}-Fig.~\ref{rsq102}):
$\eta=0.22(2),\ \delta=0.21(2)$, and $z=1.25(5)$.
Again, the exponents, except $z$ are far from the BKPZ values.
However, $\Theta$ and $z$ are consistent with the DP value.
Moreover, the sum $\eta+\delta=0.43$ that is also quite close to the
DP value (0.4732).
Thus, it is in our opinion likely that in this case the model belongs to the
DP universality class but the random synchronized states are not "natural"
absorbing states~\cite{INFINITY} and that is why $\eta$ and $\delta$
separately do not take their DP values.
Let us notice that for the Bernoulli map, examined in section III-A, the
random synchronized states are probably a good approximation of the "natural"
absorbing states, since the obtained spreading exponents coincide the DP
values.

Our numerical results are summarized in Table~\ref{tab}.
%%%%%-------------------------------------
%%%---------------------------------------
%%%%%-------------------------------------
\begin{table}
\caption{\label{tab} Critical parameters for the symmetric and asymmetric
tent map.
The last column shows how much the critcal exponents deviate from critical
hyperscaling relation (\ref{hyper}).
}
\begin{ruledtabular}
\begin{tabular} {|c|c|c|c|c|c|c|c|c|}
a   & $\gamma_c$ & $\Theta$ & $\eta$ & $\delta$ & $\beta$ & $z$
& $\eta+\delta$ & $\frac{z}{2}-\Theta-\eta-\delta$ \\\hline\hline
2 & 0.17614 & 1.26(3) & -0.53(5) & 0.0(1) & 1.50(5) & 1.30(5) &
-- & 0.07 (0.11)\\\hline
1.1 & 0.2288 & 1.16(10) & -- & -- & 1.5(1) & 1.30(5) &
-- & --\\\hline
1.02 & 0.17095 & 0.25(5) & 0.15(3) & 0.27(2) & ? & 1.32(3) & 0.42 & 0.01 (0.11)\\\hline
1.01 & 0.13695 & 0.15(2) & 0.22(2) & 0.21(2) & ? & 1.25(5) & 0.43 & 0.04 (0.08)\\\hline\hline
DP & -- & 0.1595 & 0.3137 & 0.1595 & 0.2765 & 1.2652 & 0.4732 & 0\\

\end{tabular}
\end{ruledtabular}
\end{table}
\section{Discussion}
In this paper we studied the dynamical properties of CML models
undergoing a synchronization transition.  Our main results show
that:(i) the synchronized state is not unique and (ii) non-BKPZ
critical behaviour might appear for continuous maps.  One of the open
questions that we leave for the future is why for sufficiently
asymmetric but continuous maps the linear analysis, that leads to the
relation with BKPZ model, breaks down.  For strongly asymmetric maps
the (deterministic) noise is very large and this seems to be the only
factor that might change the universality class.  It is known that
depending on how the noise scales with the order parameter,
Langevin-type models, that presumably describe our model at criticality
and at a coarsed-grained level might exhibit either DP or BKPZ critical
behaviour~\cite{MUNOZ}.  In particular when the amplitude of noise
scales linearly with the order parameter the model exhibits the BKPZ
critical behaviour while the DP one for the square root scaling.  Thus,
the problem is to explain how a qualitatively different scaling of the
noise in the Langevin-type model could emerge due to a more
quantitative changes (in asymmetricity) in a coupled CML system.  It is
also tempting to expect that BKPZ and DP critical behaviours meet for a
certain asymmetricity value $a_c$ ($1.1<a_c<1.01$) at a new
multi-critical point.  Our estimation of $\Theta$ for $a=1.02$
significantly deviates both from DP and BKPZ values which might be an
indication of such a new critical point.

\acknowledgments
This work was partially supported by the Swiss National Science Foundation
and the project OFES 00-0578 "COSYC OF SENS".
%%%%%%%%%%%%%%%%%%%%%%%%%%%%%%%%%%%%%%%%%%%%%%%%%%%%%%%%%%%%%%%%%%%%%%%%%%%%%%
%%%%%%%%%%%%%%%%%%%%%%%%%%%%%%%%%%%%%%%%%%%%%%%%%%%%%%%%%%%%%%%%%%%%%%%%%%%%%%%

%%%%%%%%%%%%%%%%%%%%%%%%%%%%%%%%%%%%%%%%%%%%%%%%%%%%%%%%%%%%%%%%%%%%%%%%%%%%%%%
%%%%%%%%%%%%%%%%%%%%%%%%%%%%%%%%%%%%%%%%%%%%%%%%%%%%%%%%%%%%%%%%%%%%%%%%%%%%%%%
\end {document}